\documentclass[12pt]{article}
\usepackage{graphicx}
\usepackage{epsf}

\makeatletter
\@addtoreset{equation}{section}
\makeatother

\def\BbbZ{Z}
\def\half{{\textstyle{1\over2}}}

\let\a=\alpha \let\b=\beta   \let\e=\epsilon

\let\la=\label  
  
 \def\bd{\begin{document}} \def\ed{\end{document}}
\def\ds{\documentstyle} \let\fr=\frac \let\bl=\bigl \let\br=\bigr
\let\Br=\Bigr \let\Bl=\Bigl
\let\bm=\bibitem
\let\na=\nabla
\let\pa=\partial \let\ov=\overline
\newcommand{\be}{\begin{equation}}
\newcommand{\ee}{\end{equation}}
\def\ba{\begin{array}}
\def\ea{\end{array}}
\newcommand{\ho}[1]{$\, ^{#1}$}
\newcommand{\hoch}[1]{$\, ^{#1}$}
\newcommand{\bea}{\begin{eqnarray}}
\newcommand{\eea}{\end{eqnarray}}
\newcommand{\ra}{\rightarrow}
\newcommand{\lra}{\longrightarrow}
\newcommand{\Lra}{\Leftrightarrow}
\newcommand{\ap}{\alpha^\prime}
\newcommand{\bp}{\tilde \beta^\prime}
\newcommand{\tr}{{\rm tr} }
\newcommand{\Tr}{{\rm Tr} }
\newcommand{\NP}{Nucl. Phys. }
\ifx\ltimes\undefined
\newcommand{\ltimes}{{\kern3pt\hbox{\vrule width 0.4pt height 5.30pt depth
.0pt}\kern-1.76pt\times\kern1pt}}
\fi

\newcommand{\tamphys}{\it $^\dag$ The Blackett Laboratory, Imperial College London,\\ 
Prince Consort Road, 
London SW7 2AZ\\
and\\
$^\ddag$Physics Department,Theory Unit, CERN, 
CH1211, Geneva23, Switzerland and INFN-Laboratori Nazionale di 
Frascati, Via E. Fermi 40, 00044 Frascati, Italy}
\newcommand{\auth}{M. J. Duff \footnote{m.duff@imperial.ac.uk}$^\dag$ and S. 
Ferrara\footnote{Sergio.Ferrara@cern.ch}$^{\ddag}$}

\begin{document}
\begin{flushright}
\hfill{Imperial/TP/2006/mjd/6}\\
\hfill{CERN-PH-TH/2006-250}\\
\hfill{hep-th/yymmnnn}\\
\end{flushright}
\vspace{10pt}

\begin{center}
{ \large {\bf Black hole entropy and quantum information
\footnote{To appear in the proceedings of the ``2006 Winter School on Attractor 
Mechanisms" (Frascati, Italy, March 2006), to be published in 
``Lecture Notes in Physics'', Springer.}
}}

\vspace{20pt}

\auth

\vspace{10pt}

{\tamphys}

\vspace{20pt}

\underline{ABSTRACT}

\end{center}

We review some recently established connections between the mathematics 
of black hole entropy in string theory and that of multipartite 
entanglement in quantum information theory. In the case of $N=2$ black 
holes and the entanglement of three qubits, the quartic $[SL(2)]^{3}$ invariant, 
Cayley's hyperdeterminant, provides both the black hole entropy and the 
measure of tripartite entanglement. In the case of $N=8$ black holes 
and the entanglement of seven qubits, the quartic $E_{7}$ invariant of Cartan 
provides both the black hole entropy and the measure of a particular 
tripartite entanglement encoded in the Fano plane. 

\vfill
\leftline{}

\newpage


\tableofcontents

\newpage


\section{Black holes and qubits}

It sometimes happens that two very different areas of theoretical physics share 
the same mathematics.  This may eventually lead to the realisation that they are, in fact,
dual descriptions of the same physical phenomena, or it may not.  Either way, it 
frequently leads to new insights in both areas. In this paper the two areas in question 
are black hole entropy in string theory and qubit entanglement in quantum information 
theory. Going one way, we shall learn that the entropy of the so-called 
STU $N=2$ black hole is given by the ``hyperdeterminant'', a quantity first introduced by 
Cayley in 1845 and which describes the tripartite entanglement of three 
qubits \cite{Duff:2006uz,Kallosh:2006zs,Levay:2006kf}.  Going the other way, we discover that the exceptional 
group $E_{7}$, the U-duality group of $N=8$ supergravity, plays a part in the 
tripartite entanglement of seven qubits \cite{Duff:2006ue,Levay:2006pt}. 

We begin in section \ref{STU} with an interesting subsector of string compactification to four 
dimensions which is provided by the $STU$ model whose low energy limit is 
described by $N=2$ supergravity coupled to three vector multiplets. 
One may regard it as a truncation of an $N=4$ 
theory obtained by compactifying the heterotic string on 
$T^{6}$ where $S,T,U$ correspond to the dilaton/axion, complex Kahler form 
and complex structure fields respectively. It exhibits an $SL(2,\BbbZ)_S$ 
strong/weak coupling duality 
and an $SL(2,\BbbZ)_T \times SL(2,\BbbZ)_U$ target space duality. 
By string/string duality, 
this is equivalent to a Type IIA string on $K3 \times T^{2}$ with $S$ and 
$T$ exchanging roles \cite{Duff:1993ij,Hull:1994ys,Duff:1994zt}. Moreover, by mirror symmetry 
 this is in turn equivalent to a Type IIB string on the mirror manifold with $T$ and 
$U$ exchanging roles. Another way to obtain this model is by 
truncation of the N=8 theory that results from $T^{7}$ 
compactification of M-theory. Either way, the truncated theory has a combined 
$[SL(2,Z)]^{3}$ duality and complete $S-T-U$ triality symmetry \cite{Duff:1995sm}.
Alternatively, one may simply start with this $N=2$ theory directly as 
an interesting four-dimensional supergravity in its own right, as 
described in section \ref{STU}. 

The model admits extremal black holes solutions carrying four electric and 
magnetic charges and  we organize these 8 
charges into the  $2 \times 2 \times 2$ {\it hypermatrix}, $a_{ABD}$, and display 
the $S-T-U$ symmetric Bogomolnyi mass formula \cite{Duff:1995sm}. 
Associated with this hypermatrix is a {\it hyperdeterminant}, ${\rm 
Det}~a_{ABD}$, 
discussed in section \ref{Cayley}, first introduced
by Cayley in 1845 \cite{Cayley}.  The black hole entropy, first calculated in \cite{Behrndt:1996hu}, 
is quartic in the charges and must be invariant under $[SL(2,Z)]^{3}$ and under triality. The main 
result of section \ref{Black}, is to 
show \cite{Duff:2006uz} that this entropy given by the square root 
of Cayley's hyperdeterminant:

\be
 S= \pi \sqrt{|{\rm Det}~a_{ABD}|}.
\ee

The hyperdeterminant also makes it appearance in quantum information 
theory \cite{Miyake}.  Let the three qubit system $ABD$ (Alice, Bob 
amd Daisy) be in a pure state $|\Psi\rangle$, 
and let the components of $|\Psi\rangle$ in the standard basis be 
$a_{ABD}$:
\begin{equation}
|\Psi\rangle = a_{ABD}|ABD\rangle
\end{equation}
or
\[
|\Psi\rangle = a_{000}|000\rangle+a_{001}|001\rangle+a_{010}|010\rangle+a_{011}|011\rangle
\]
\be
+a_{100}|100\rangle+a_{101}|101\rangle+a_{110}|110\rangle+a_{111}|111\rangle
\ee
Then the three way entanglement of the three qubits $A$, $B$ and 
$D$ is given by the {\it 3-tangle} \cite{Coffman:1999jd}
\be
\tau_{3}(ABD)=4 |{\rm Det}~a_{ABD}|.
\ee
The 3-tangle is maximal for the GHZ state $|000\rangle+|111\rangle$ 
\cite{Greenberger} and vanishes for the states $p|100\rangle+q|010\rangle+r|001\rangle$.
The relation between three qubit quantum entanglement and the Cayley 
hyperdeterminant was pointed out by Miyake and Wadati \cite{Miyake}.

As far as we can tell \cite{Duff:2006uz}, the appearance of the Cayley 
hyperdeterminant in these two different contexts of stringy black hole entropy
(where the $a_{ABD}$ are integers and the symmetry is 
$[SL(2,Z)]^{3}$) and three-qubit quantum entanglement (where the $a_{ABD}$ 
are complex numbers and the symmetry is 
$[SL(2,C]^{3}$) is a purely 
mathematical coincidence. Nevertheless, it has already provided 
fascinating new insights 
\cite{Duff:2006uz,Kallosh:2006zs,Levay:2006kf,Duff:2006ue,Levay:2006pt} into the connections between strings, black holes, and 
quantum information\footnote{A third 
application \cite{Duff:2006ev}, not considered in this paper, is the Nambu-Goto string whose 
action is also given by $\sqrt{|{\rm Det}~a_{ABD}|}$ in spacetime signature 
$(2,2)$.}.

In section \ref{N=8} we extend the argument to the $N=8$ case and, noting that
\be
E_{7(7)}(Z) \supset [SL(2,Z)]^{7}
\ee
and  
\be
E_{7}(C)\supset [SL(2,C)]^{7},
\ee
show that the corresponding system in quantum information theory is that 
of seven qubits (Alice, Bob, Charlie, Daisy, Emma, Fred and George). However, the larger symmetry requires that they undergo at most tripartite 
entanglement of a very specific kind. As discussed in section 
\ref{Cartan}, the entanglement measure will be given by the quartic 
Cartan $E_{7}(C)$ invariant \cite{Cartan,Cremmer:1979up,Kallosh:1996uy,Ferrara:1997uz}. The 
entanglement may be represented by the Fano plane \cite{Klein} which also 
provides the multiplication table of the  octonions.  See also the interesting 
paper by Levay \cite{Levay:2006pt} who noted independently the connection to 
the Fano plane.

\section{The N=2 STU model}
\la{STU}

Consider the three complex scalars axion/dilaton
field $S$, the complex Kahler form field $T$ and the complex structure
field $U$
\begin{eqnarray}
S&=&S_1+iS_2\nonumber\\
T&=&T_1+iT_2\nonumber\\
U&=&U_1+iU_2\ .
\end{eqnarray}
This complex parameterization allows for a natural transformation under the
various $SL(2,\BbbZ)$ symmetries.  The action of $SL(2,\BbbZ)_S$ is given by
\be
S \rightarrow \frac{aS+b}{cS+d}\ ,
\la{sl2zs}
\ee
where $a,b,c,d$ are integers satisfying $ad-bc=1$, with similar
expressions for $SL(2,\BbbZ)_T$ and $SL(2,\BbbZ)_U$.  Defining the
matrices ${\cal M}_S$, ${\cal M}_T$ and ${\cal M}_U$ via
\be
{\cal M}_S=\frac{1}{S_2}
\left(
\begin{array}{cc}
1 & S_1\\
S_1 & |S|^2
\end{array}
\right)\ ,
\label{eq:sl2mat}
\ee
the action of $SL(2,\BbbZ)_S$ now takes the form
\be
{\cal M}_S\rightarrow \omega_S{}^T{\cal M}_S\omega_S\ ,
\ee
where
\be
\omega_S=
\left(
\begin{array}{cc}
d& b\\
c & a
\end{array}
\right)\ ,
\ee
with similar expressions for ${\cal M}_T$ and ${\cal M}_U$.
We also define the $SL(2,\BbbZ)$ invariant tensors
\be
\epsilon_S=\epsilon_T=\epsilon_U=
\left(
\begin{array}{cc}
0& 1\\
-1 & 0
\end{array}
\right)\ .
\ee

Starting from the heterotic string,the bosonic action for the graviton $g_{\mu\nu}$, dilaton 
$\eta$, two-form $B_{\mu\nu}$ four $U(1)$ gauge fields $A_S^a$
and two complex scalars $T$ and $U$ is \cite{Duff:1995sm}
\begin{eqnarray}
I_{STU}&=&\frac{1}{16\pi G}\int d^4x\sqrt{-g}e^{-\eta}\Bigl[
R_g + g^{\mu\nu}\partial_{\mu}\eta\partial_{\nu}\eta
-\frac{1}{12}g^{\mu\lambda}g^{\nu\tau}g^{\rho\sigma}
H_{\mu\nu\rho}H_{\lambda\tau\sigma}\nonumber\\
&&\kern9.3em
+\frac{1}{4}\Tr(\partial{\cal M}_T{}^{-1}\partial {\cal M}_T)
+\frac{1}{4}\Tr(\partial{\cal M}_U{}^{-1}\partial {\cal M}_U)\nonumber\\
&&\kern9.3em
-\frac{1}{4}{F_S}_{\mu\nu}{}^T({\cal M}_T \times {\cal M}_U){F_S}^{\mu\nu}
\Bigr]\ .
\la{S}
\end{eqnarray}
where the metric $g_{\mu\nu}$ is related to the four-dimensional canonical 
Einstein metric 
$g^c_{\mu\nu}$ by $g_{\mu\nu}=e^{\eta}g^c{}_{\mu\nu}$ 
and where
\be
H_{\mu\nu\rho}=3(\partial_{[\mu}B_{\nu\rho]}
-\half A_{S[\mu}{}^T (\epsilon_T\times\epsilon_U){F_S}_{\nu\rho]}).
\ee
This action is
manifestly invariant under $T$-duality and $U$-duality, with
\be
{F_S}_{\mu\nu}\rightarrow
(\omega_T{}^{-1}\times\omega_U{}^{-1}){F_S}_{\mu\nu}\ , \, \qquad
{\cal M}_{T/U}\rightarrow \omega_{T/U}^T \, {\cal M}_{T/U} \, \omega_{T/U}\ ,
\ee
and with $\eta$, $g_{\mu\nu}$ and $B_{\mu\nu}$ inert.  Its equations of
motion and Bianchi identities (but not the action itself) are also
invariant under $S$-duality (\ref{sl2zs}), with $T$ and $g^c{}_{\mu\nu}$ inert and
with
\be
\left(
\begin{array}{c}
{{F_S}}_{\mu\nu}{}^a\\
{\widetilde{F}_S}{}_{\mu\nu}{}^a
\end{array}
\right)
\rightarrow     \omega_S^{-1}
\left(
\begin{array}{c}
{{F_S}}_{\mu\nu}{}^a\\
{\widetilde{F}_S}{}_{\mu\nu}{}^a
\end{array}
\right)\ ,
\ee
where
\be
{\widetilde{F}_S}{}_{\mu\nu}{}^{a}=-S_2[({\cal M}_T{}^{-1} \times {\cal
M}_U{}^{-1})(\epsilon_T \times \epsilon_U)]^a{}_b  *
{F_S}_{\mu\nu}{}^{b}-S_1 {F_S}_{\mu\nu}{}^{a}  \ ,
\ee
where the axion field $a$ is defined by
\be
\epsilon^{\mu\nu\rho\sigma}\partial_{\sigma}a=
\sqrt{-g}e^{-\eta}g^{\mu\sigma}g^{\nu\lambda}g^{\rho\tau}
H_{\sigma\lambda\tau}\ ,
\ee
and where $S=S_1+iS_2=a+ie^{-\eta}$.

Thus $T$-duality transforms Kaluza-Klein electric charges
$({F_S}^3,{F_S}^4)$ into winding electric charges $({F_S}^1,{F_S}^2)$
(and Kaluza-Klein magnetic charges into winding magnetic charges),
$U$-duality transforms the Kaluza-Klein and winding electric charge of
one circle $({F_S}^3,{F_S}^2)$ into those of the other
$({F_S}^4,{F_S}^1)$ (and similarly for the magnetic charges) but
$S$-duality transforms Kaluza-Klein electric charge $({F_S}^3,{F_S}^4)$
into winding magnetic charge $({\tilde {F_S}}^3,{\tilde {F_S}}^4)$ (and
winding electric charge into Kaluza-Klein magnetic charge). In summary
we have $SL(2,\BbbZ)_T \times SL(2,\BbbZ)_U$ and $T \leftrightarrow U$
off-shell but $SL(2,\BbbZ)_S \times SL(2,\BbbZ)_T \times SL(2,\BbbZ)_U$
and an $S$--$T$--$U$ interchange on-shell.  

One may also consider the Type IIA action  $I_{TUS}$ and the Type IIB action $I_{UST}$ obtained by 
cyclic permutation of the fields $S,T,U$. Finally, one may consider an action \cite{Behrndt:1996hu} where the $S$, $T$ and $U$ fields 
enter democratically with a prepotential
\be
F=STU
\label{eq:Haxion}
\ee
which off-shell has the full $STU$ interchange but none of the $SL(2,Z)$. All four versions are 
on-shell equivalent.

Following \cite{Duff:1995sm}, it is now straightforward to write down an 
$S$--$T$--$U$ symmetric
Bogomolnyi mass formula. Let us define electric and magnetic charge
vectors $\alpha_S^a$ and $\beta_S^a$ associated with the field strengths
${{F_S}}^a$ and ${\tilde {F_S}}^a$ in the standard way.
The electric and magnetic charges $Q_S^a$ and $P_S^a$ are
given by
\be {F_{S}}_{0r}^a\sim\frac{Q_S^a}{r^2} \, \qquad
*{F_{S}}_{0r}^a\sim\frac{P_S^a}{r^2}\ ,
\ee
giving rise to the charge vectors
\be
\pmatrix{\a_S^a\cr \b_S^a}=\pmatrix{  S_2^{(0)} {\cal M}_T^{-1}
   \times {\cal M}_U^{-1} & S_1^{(0)} \e_T \times \e_U  \cr 0 &
-\e_T \times \e_U }^{ab} \pmatrix{Q_S^b \cr P_S^b}.
\ee
For our purpose it is useful to define a $2 \times 2 \times 2$ array
 $a_{ABD}$ via
\be
\left(
\begin{array}{c}
a_{000}\\
a_{001}\\
a_{010}\\
a_{011}\\
a_{100}\\
a_{101}\\
a_{110}\\
a_{111}
\end{array}
\right)
=
\left(
\begin{array}{c}
-\beta_S^1\\
-\beta_S^2\\
-\beta_S^3\\
-\beta_S^4\\
\alpha_S^1\\
\alpha_S^2\\
\alpha_S^3\\
\alpha_S^4
\end{array}
\right)\ ,
\ee
transforming as
\be
a^{ABD}\rightarrow
\omega_S{}^{A}{}_{A'}
\omega_T{}^{B}{}_{B'}
\omega_U{}^{D}{}_{D'}
a^{A'B'D'}\ .
\ee
Then the mass formula is
\begin{equation}
m^2=\frac{1}{16}a^T({\cal M}_S{}^{-1}{\cal M}_T{}^{-1}{\cal M}_U{}^{-1}
		-{\cal M}_S{}^{-1}{\epsilon}_T{\epsilon}_U
		-{\epsilon}_S{\cal M}_T{}^{-1}{\epsilon}_U
		-{\epsilon}_S{\epsilon}_T{\cal M}_U{}^{-1})a\ .
\la{us}
\end{equation}
This is consistent with the general $N=2$ Bogomolnyi formula \cite{Ceresole:1994cx}. Although all theories 
have the same mass spectrum, there is clearly a difference of interpretation with electrically charged
elementary states in one picture being solitonic monopole or dyon
states in the other. 

This $2 \times 2 \times 2$ array $a_{ABD}$ is an example  a ``hypermatrix'', a term 
coined by Cayley in 1845 \cite{Cayley} where he also introduced a ``hyperdeterminant''.

\section{Cayley's hyperdeterminant}
\la{Cayley}

In 1845 Cayley \cite{Cayley} generalized the determinant of a $2 \times 2$ matrix 
$a_{AB}$ to the {\it hyperdeterminant} of a $2 \times 2 \times 2$ {\it hypermatrix} 
$a_{ABD}$
\[
{\rm Det}~a=-\frac{1}{2}
\epsilon^{A_{1}A_{2}}\epsilon^{B_{1}B_{2}}\epsilon^{D_{1}D_{4}}\epsilon^{A_{3}A_{4}}\epsilon^{B_{3}B_{4}}\epsilon^{D_{2}D _{3}}
{a}_{A_{1}B_{1}D_{1}}{a}_{A_{2}B_{2}D_{2}}{a}_{A_{3}B_{3}D_{3}}{a}_{A_{4}B_{4}D_{4}}
\]
\[
=  a_{000}^2 a_{111}^2 + a_{001}^2 a_{110}^2 +
        a_{010}^2 a_{101}^2 + a_{100}^2 a_{011}^2 
\]
\[
        -2(a_{000}a_{001}a_{110}a_{111}+a_{000}a_{010}a_{101}a_{111}
\]
\[
        + a_{000}a_{100}a_{011}a_{111}+a_{001}a_{010}a_{101}a_{110}
\]
\[
        + a_{001}a_{100}a_{011}a_{110}+a_{010}a_{100}a_{011}a_{101}) 
\]
\be
        + 4 (a_{000}a_{011}a_{101}a_{110} + a_{001}a_{010}a_{100}a_{111})
\la{hyperdet}
\ee
\[
  =a_{0}^2 a_{7}^2 + a_{1}^2 a_{6}^2 +
        a_{2}^2 a_{5}^2 + a_{3}^2 a_{4}^2 
\]
\[
        -2(a_{0}a_{1}a_{6}a_{7}+a_{0}a_{2}a_{5}a_{7}
+ a_{0}a_{4}a_{3}a_{7}+a_{1}a_{2}a_{5}a_{6}
+ a_{1}a_{3}a_{4}a_{6}+a_{2}a_{3}a_{4}a_{5}) 
\]
\be
        + 4 (a_{0}a_{3}a_{5}a_{6} + a_{1}a_{2}a_{4}a_{7})
\la{hyperdet1}
\ee
where we have made the binary conversion $0, 1, 2, 3, 4, 5, 6, 7$ for 
$000, 001, 010, 011, 100, 101, 110, 111$.

The hyperdeterminant vanishes iff the following system of equations 
in six unknowns $p^{A},q^{B},r^{D}$ has a nontrivial solution, not 
allowing any of the pairs to be both zero:
\[
a_{ABD}p^{A}q^{B}=0
\]
\[
a_{ABD}p^{A}r^{D}=0
\]
\be
a_{ABD}q^{B}r^{D}=0
\ee
For our purposes, the important properties of the hyperdeterminant are that it is a quartic 
invariant under $[SL(2)]^{3}$ and under a triality that interchanges 
$A$, $B$ and $D$. These properties are valid whether the $a_{ABD}$ 
are complex, real or integer.

One way to understand this triality is to think of  
having  three different metrics (Alice, Bob and Daisy) 
\[
(\gamma_{A})_{A_{1}A_{2}}=\epsilon^{B_{1}B_{2}}\epsilon^{D_{1}D_{2}}a_{A_{1}B_{1}D_{1}}a_{A_{2}B_{2}D_{2}}
\]
\[
(\gamma_{B})_{B_{1}B_{2}}=\epsilon^{D_{1}D_{2}}\epsilon^{A_{1}A_{2}}a_{A_{1}B_{1}D_{1}}a_{A_{2}B_{2}D_{2}}
\]
\be
(\gamma_{D})_{D_{1}D_{2}}=\epsilon^{A_{1}A_{2}}\epsilon^{B_{1}B_{2}}a_{A_{1}B_{1}D_{1}}a_{A_{2}B_{2}D_{2}}
\ee
Explicitly,
\be
{\gamma}= 
\pmatrix{2(a_{0}a_{6}-a_{2}a_{4}) &  a_{0}a_{7}-a_{2}a_{5}+a_{1}a_{6}-a_{3}a_{4}\cr 
        a_{0}a_{7}-a_{2}a_{5}+a_{1}a_{6}-a_{3}a_{4}  & 2(a_{1}a_{7}-a_{3}a_{5})}
\ee
\be
{\beta}= 
\pmatrix{2(a_{0}a_{3}-a_{1}a_{2}) &  a_{0}a_{7}-a_{1}a_{6}+a_{4}a_{3}-a_{5}a_{2}\cr 
        a_{0}a_{7}-a_{1}a_{6}+a_{4}a_{3}-a_{5}a_{2}  & 2(a_{4}a_{7}-a_{5}a_{6})}
\ee
\be
{\alpha}= 
\pmatrix{2(a_{0}a_{5}-a_{4}a_{1}) &  
a_{0}a_{7}-a_{4}a_{3}+a_{2}a_{5}-a_{6}a_{1}\cr 
        a_{0}a_{7}-a_{4}a_{3}+a_{2}a_{5}-a_{6}a_{1}  & 
        2(a_{2}a_{7}-a_{6}a_{3})}
\ee
All are equivalent, however, since
\be
det~\alpha=det~\beta=det~\gamma=-{\rm Det}~a
\ee
If we make the identifications

\bea\label{dictionary}
a_{0}~ =&   \frac{1}{\sqrt{2}}(-P^{0}+P^{2})\nonumber\\
a_{1}~ =&   \frac{1}{\sqrt{2}}(-Q^{0}+Q^{2})\nonumber\\
a_{2}~ =&   \frac{1}{\sqrt{2}}(~P^{1}-P^{3})\nonumber\\
a_{3}~ =&   \frac{1}{\sqrt{2}}(~Q^{1}-Q^{3})\nonumber\\
a_{4}~ =&   \frac{1}{\sqrt{2}}(-P^{1}-P^{3})\nonumber\\
a_{5}~ =&   \frac{1}{\sqrt{2}}(-Q^{1}-Q^{3})\nonumber\\
a_{6}~ =&   \frac{1}{\sqrt{2}}(-P^{0}-P^{2})\nonumber\\
a_{7}~ =&   \frac{1}{\sqrt{2}}(-Q^{0}-Q^{2})
\la{charges}
\end{eqnarray}
then we find the $O(2,2)$ scalar products 
\[
2(a_{0}a_{6}-a_{2}a_{4})=( P^0)^2 + ( P^1)^2 -( P^2)^2- ( P^3)^2=P^{2}
\]
\[
2(a_{1}a_{7}-a_{3}a_{5})=( Q_0)^2 + ( Q_1)^2 -( Q_2)^2- ( Q_3)^2=Q^{2}
\]
\[
a_{0}a_{7}-a_{2}a_{5}+a_{1}a_{6}-a_{3}a_{4}=( P^0 Q_0) + ( P^1 Q_1)  +( P^2 Q_2)+ ( P^3 Q_3)=P.Q
\]
so
\be
{\gamma}= 
\pmatrix{ P^{2}& P.Q \cr 
          P.Q  & Q^{2}}
\ee
and
\[
-Det~a=P^{2}Q^{2}-(P.Q)^{2}
\]

\section{Black hole entropy}
\la{Black}
The $STU$ model admits extremal black hole solutions satisfying the 
Bogomolnyi mass formula. As usual, their entropy is given by one 
quarter the area of the event horizon. However, to calculate this 
area requires evaluating the mass not with the asymptotic values of the 
moduli, but with their frozen values on the horizon which are fixed 
in terms of the charges \cite{Ferrara:1995ih}. This ensures that the entropy is 
moduli-independent, as it should be. The relevant calculation was 
carried out in \cite{Behrndt:1996hu} for the model with the $STU$ 
prepotential. The electric and magnetic charges of that paper are denoted
$(p^0, q_0),\, (p^1, q_1),\, (p^2, q_2),\, (p^3,q_3)$.  In these variables, the entropy is given by
\be
S=\pi \left( W(p^\Lambda,q_\Lambda)\right)^{1/2}
\ee
where
\begin{equation} \label{ww}
 W(p^\Lambda ,q_\Lambda) =-{(p\cdot q)}^2+4\bigl (
(p^1q_1)(p^2q_2)+(p^1q_1)(p^3q_3)+(p^3q_3)(p^2q_2)\bigr )\\
 - 4 p^0 q_1 q_2 q_3 + 4q_0 p^1 p^2 p^3 \ .
\end{equation}
The function $ W(p^\Lambda ,q_\Lambda)$ is symmetric under
transformations: $
p^1\leftrightarrow p^2 \leftrightarrow p^3 $
and  $ q_1\leftrightarrow q_2 \leftrightarrow q_3. $
For the solution to be BPS we have to require $ W>0$.

If we make the identifications \cite{Duff:2006uz}
\begin{equation}
\left [\matrix{
 p^0\cr
 p^1\cr
 p^2\cr
 p^3\cr
 q_0\cr
 q_1\cr
 q_2\cr
 q_3\cr
}\right ]= 
 \left [\matrix{
-a_{0}\cr
-a_{1}\cr
-a_{2}\cr
 a_{4}\cr
-a_{7}\cr
 a_{6}\cr
 a_{5}\cr
-a_{3}\cr
}\right ]
\label{charges7}\end{equation}
we recognize from (\ref{Cayley}) that
\be
W=-{\rm Det}~a_{}
\ee
and hence the black hole entropy is given by
\be
S=\pi \sqrt{ -{\rm Det}~a_{}}
\ee

Some examples of supersymmetric black hole solutions \cite{Duff:1994jr} are provided by the 
electric Kaluza-Klein black hole with $\alpha=(1,0,0,0)$ and $\beta=(0,0,0,0)$; the electric 
winding black hole with $\alpha=(0,0,0,-1)$ and $\beta=(0,0,0,0)$; 
the magnetic Kaluza-Klein black hole with $\alpha=(0,0,0,0)$ and 
$\beta=(0,-1,0,0)$; the magnetic winding black hole with 
$\alpha=(0,0,0,0)$ and $\beta=(0,0,-1,0)$. These are characterized by a 
scalar-Maxwell coupling parameter $a=\sqrt{3}$. By combining these 1-particle states, we may build up 2-, 3- and 4-particle bound states at threshold 
\cite{Duff:1994jr,Duff:1995sm}. For example $\a=(1,0,0,-1)$ and $\beta=(0,0,0,0)$ with $a=1$; 
$\a=(1,0,0,-1)$ and $\b=(0,-1,0,0)$ with $a= 1/\sqrt{3}$; $\alpha=(1,0,0,-1)$ and 
$\beta=(0,-1,-1,0)$ with $a=0$. The 1-, 2- and 3-particle states all yield 
vanishing contributions to ${\rm Det}~a_{}$. A non-zero value is obtained for the 
4-particle example, however, which is just the Reissner-Nordstrom black 
hole. 

\section{The N=8 generalization}

The black holes described by Cayley's hyperdeterminant are those of 
$N=2$ supergravity coupled to three vector multiplets, where the 
symmetry is $[SL(2,Z)]^{3}$.  One might therefore ask whether the black hole/information 
theory correspondence could be generalized. There are three 
generalizations we might consider:

1) $N=2$ supergravity coupled to $l$ vector multiplets where the 
symmetry is $SL(2,Z) \times SO(l-1,2,Z)$ and the black holes carry charges 
belonging to the $(2,l+1)$ representation ($l+1$ electric plus $l+1$ magnetic).

2) $N=4$ supergravity coupled to $m$ vector multiplets where the 
symmetry is $SL(2,Z) \times SO(6,6+m,Z)$ where the black holes carry 
charges belonging to the $(2,12+m)$ representation ($m+12$ electric plus 
$m+12$ magnetic). 

3) $N=8$ supergravity where the symmetry is the non-compact 
exceptional group $E_{7(7)}(Z)$ and the black 
holes carry charges belonging to the fundamental $56$-dimensional 
representation (28 electric plus 28 magnetic).

In all three case there exit quartic invariants akin to Cayley's 
hyperdeterminant whose square root yields the corresponding black hole 
entropy. If there is to be a quantum information theoretic 
interpretation, however, it cannot just be 
random entanglement of more qubits, because the general $n$ qubit entanglement is 
described by the group $[SL(2,C)]^{n}$, which, even after replacing $Z$ by $C$, differs from the 
above symmetries (except when $n=3$, which correspond to case (1) 
above with $l=3$, the case we already know.).

We note, however, that
\be
E_{7(7)}(Z) \supset [SL(2,Z)]^{7}
\ee
and  
\be
E_{7}(C)\supset [SL(2,C)]^{7},
\ee
We shall now show that the corresponding system in quantum information theory is that 
of seven qubits (Alice, Bob, Charlie, Daisy, Emma, Fred and George). However, the larger symmetry requires that they undergo at most tripartite 
entanglement of a very specific kind. The entanglement measure will be given by the quartic 
Cartan $E_{7}(C)$ invariant \cite{Cartan,Cremmer:1979up,Kallosh:1996uy,Ferrara:1997uz}.

\section{Decomposition of $E_{7(7)}$}
\la{N=8}

Consider the decomposition of the fundamental 56-dimensional 
representation of $E_{7(7)}$ under its maximal subgroup

\[
E_{7(7)} \supset  SL(2)_{A} \times SO(6,6)
\]
\be
56 \rightarrow (2,12) + (1,32)
\la{56}
\ee
Further decomposing $SO(6,6)$, 
\[
SL(2)_{A} \times SO(6,6) \supset SL(2)_{A} \times SL(2)_{B} \times 
SL(2)_{D} \times SO(4,4) 
\]
\[
(2,12)+(1,32)  \rightarrow  (2,2,2,1)
\]
\be
+(2,1,1,8_{v}) + (1,2,1,8_{s}) +(1,1,2,8_{c})
\la{triality}
\ee
Further decomposing $SO(4,4)$,
\[
SL(2)_{A} \times SL(2)_{B} \times SL(2)_{D} \times SO(4,4) \supset 
SL(2)_{A} \times SL(2)_{B} \times SL(2)_{D}
\]
\[
\times SO(2,2) \times SO(2,2)
\]
\[
(2,2,2,1)+ (2,1,1,8_{v}) + (1,2,1,8_{s}) +(1,1,2,8_{c}) \rightarrow
\]
\[
(2,2,2,1,1)+ (2,1,1,4,1) +(2,1,1,1,4) 
\]
\be 
+(1,2,1,2,2) + (1,2,1,2,2) + (1,1,2,2,2) +(1,1,2,2,2)
\ee
Finally, further decomposing each $SO(2,2)$
\[
SL(2)_{A} \times SL(2)_{B} \times SL(2)_{D} \times SO(2,2) \times SO(2,2) 
\supset
\]
\[
SL(2)_{A} \times SL(2)_{B} \times SL(2)_{D} \times SL(2)_{C} \times SL(2)_{G}
 \times SL(2)_{F} \times SL(2)_{E}
\]
\[
(2,2,2,1,1) + (2,1,1,4,1) +(2,1,1,1,4) 
\]
\[
+(1,2,1,2,2) + (1,2,1,2,2) + (1,1,2,2,2) +(1,1,2,2,2) 
\rightarrow
\]
\[
(2,2,2,1,1,1,1) + (2,1,1,2,2,1,1) + (2,1,1,1,1,2,2) +  
\]
\[
(1,2,1,2,1,1,2) +(1,2,1,1,2,2,1) + (1,1,2,2,1,2,1) + (1,1,2,1,2,1,2)
\]

In summary, 
\be
E_{7(7)} \supset 
SL(2)_{A} \times
SL(2)_{B} \times
SL(2)_{C} \times
SL(2)_{D} \times
SL(2)_{E} \times
SL(2)_{F} \times
SL(2)_{G} 
\ee
and the 56 decomposes as
\[
56 
\rightarrow
\]
\[
~(2,2,1,2,1,1,1)  
\]
\[ 
+(1,2,2,1,2,1,1)
\]
\[
+(1,1,2,2,1,2,1)      
\]
\[
+(1,1,1,2,2,1,2)              
\]
\[
+(2,1,1,1,2,2,1)                                            
\]
\[ 
+(1,2,1,1,1,2,2) 
\]
\be
 +(2,1,2,1,1,1,2)
\la{decomp}
\ee

An analogous decomposition holds for 
\be
E_{7}(C)\supset [SL(2,C)]^{7}.
\ee

\section{Tripartite entanglement of 7 qubits}

We have seen that in the case of three qubits, the tripartite 
entanglement is described by $[SL(2,C)]^{3}$ and that the entanglement 
measure is given by Cayley's hyperdeterminant.  Now we consider seven 
qubits (Alice, Bob, Charlie, Daisy, Emma, Fred and George) but where 
Alice has tripartite entanglement not only with Bob/Daisy but also with 
Emma/Fred and also with George/Charlie, and similarly for the other six 
individuals. So, in fact, each person has tripartite 
entanglement with each of the remaining three couples:
\[
|\Psi\rangle = 
\]
\[
                a_{ABD}|ABD\rangle
                \]
                \[
               +b_{BCE}|BCE\rangle
               \]
               \[
               +c_{CDF}|CDF\rangle
               \]
               \[
               +d_{DEG}|DEG\rangle
               \]
               \[
               +e_{EFA}|EFA\rangle
               \]
               \[
               +f_{FGB}|FGB\rangle
               \]
\be
               +g_{GAC}|GAC\rangle
			   \la{psi}
\ee

Note that:

1) Any pair of states has an individual in common

2) Each individual is excluded from four out of the seven states

3) Two given individuals are excluded from two out of the seven states

4) Three given individuals are never excluded

The entanglement may be represented by a heptagon with vertices 
A,B,C,D,E,F,G and seven triangles ABD, BCE, CDF, DEG, EFA, FGB, and 
GAC. See Figure 1.
Alternatively, we can use the Fano plane. See Figure 2.
The Fano plane corresponds to the multiplication table of the  
octonions as may be seen from the description of the state $|\Psi\rangle$ 
given in Table 1.

\begin{figure}[ht]
\begin{center}
\epsfysize=7cm\epsfbox{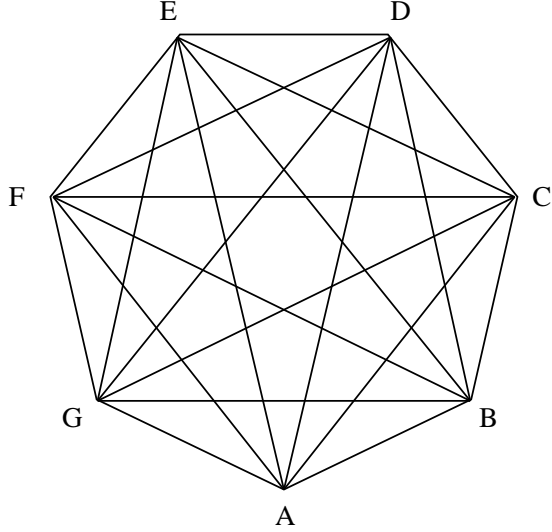}
\end{center}
\caption{The $E_7$ entanglement diagram. Each of the seven vertices A,B,C,D,E,F,G 
represents a qubit and each of the seven triangles ABD, BCE, CDF, DEG, 
EFA, FGB, GAC describes a tripartite entanglement.  }
\end{figure} 

\begin{figure}[ht]
\begin{center}
\epsfysize=7cm\epsfbox{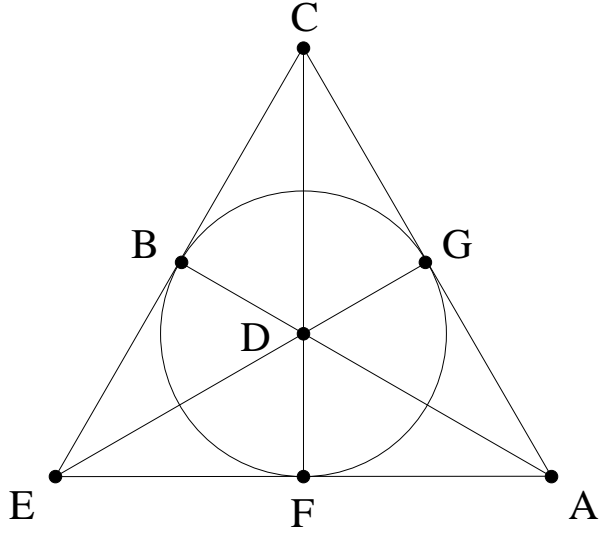}
\end{center}
\caption{The Fano plane has seven points, representing the seven 
qubits, and seven lines (the circle 
counts as a line) with three points on every line, representing the 
tripartite entanglement, and three lines through every point.}
\end{figure}

\begin{table}
\begin{center}
\renewcommand{\arraystretch}{1.6}
\renewcommand{\tabcolsep}{4pt}
\begin{tabular}{c|ccccccc}
\hline
~&A&B&C&D&E&F&G\\
\hline
A &  & D & G &-B & F &-E &-C \\
B &-D &  & E & A & -C& G &-F \\
C &-G &-E &  & F & B &-D & A \\
D & B &-A &-F &  & G & C &-E \\
E &-F & C &-B &-G &  & A & D \\
F & E &-G & D &-C &-A &  & B \\
G & C & F&-A  & E &-D &-B & \\
\hline
\end{tabular}
\caption{The entanglement of the state $|\Psi\rangle$ coincides with the 
multiplication table of the  octonions.}
\label{octonions}
\end{center}
\end{table}

Each of the seven states transforms as a $(2,2,2)$ under three of the 
$SL(2)$'s and are singlets under the remaining four. Note that from 
(\ref{triality}) we see that the A-B-C triality of 
section \ref{Cayley} is linked with the $8_{v}-8_{s}-8_{c}$ triality 
of the $SO(4,4)$. For example, interchanging A and B leaves 
$|\Psi\rangle$ invariant provided we also interchange C and F. 
Individually, therefore, the tripartite entanglement of each of 
the seven states is given by Cayley's 
hyperdeterminant. Taken together however, we see from (\ref{decomp}) 
that they transform as a complex $56$ of $E_{7}(C)$. Their 
tripartite entanglement must be is given by an expression that is quartic 
in the coefficients $a,b,c,d,e,f,g$ and invariant under $E_{7}(C)$.
The unique possibility is the Cartan invariant $I_{4}$, and so the 3-tangle is 
given by

\be
\tau_{3}(ABCDEFG) = 4 |I_{4}|
\ee
If the wave-function (\ref{psi}) is normalized, then $0 \leq \tau_{3}(ABCDEFG) 
\leq 1$.
\section{Cartan's $E_{7(7)}$ invariant}
\la{Cartan}

The Cremmer-Julia \cite{Cremmer:1979up} form of the Cartan $E_{7(7)}$ invariant 
may be written as 
\begin{eqnarray}
I_{4} &=&  {\mbox Tr} (Z\bar Z )^2 -{\textstyle{1\over 4}}
({\mbox Tr}\,Z\bar Z )^2 + 4  ({\mbox P\hskip- .1cm f}\; Z +
{\mbox P\hskip- .1cm f} \; \bar Z \,)  \ ,
\label{diamond}
\end{eqnarray}
and the Cartan form \cite{Cartan} may be written as 
\begin{eqnarray}
I_{4}&=&-{\mbox Tr} (\, x \; y )^2 + {\textstyle{1\over 4}}
({\mbox Tr}\, x \;   y)^2 - 4  ({\mbox P\hskip- .1cm f}~ x +
{\mbox P\hskip- .1cm f} ~ y \,)  \ .
\label{cartan}\end{eqnarray}
Here
\begin{equation}
Z_{AB} = -{1\over 4\sqrt 2}(x^{ab} + i y_{ab})(\Gamma^{ab})_{AB}
\label{Z}\end{equation}
and
\begin{equation}
x^{ab} + i y_{ab} = -{\sqrt 2\over 4} Z_{AB}  (\Gamma^{AB})_{ab}
\label{xy}\end{equation}
The matrices of the $SO(8)$ algebra are $(\Gamma^{ab})_{AB}$ where $(a \, 
b)$ are the 8 vector indices and $(A,B)$ are the 8 spinor indices. 
The $(\Gamma^{ab})_{AB}$ matrices can be considered also as 
$(\Gamma^{AB})_{ab}$ matrices due to equivalence of the vector and 
spinor representations of the $SO(8)$ Lie algebra. 
The exact relation between the Cartan invariant  in 
(\ref{cartan})  and Cremmer-Julia  invariant \cite{Cremmer:1979up} in 
(\ref{diamond}) was established in 
\cite{Balasubramanian:1997az,Gunaydin}. The quartic invariant $I_4$ of $E_{7(7)}$ 
is also related to the octonionic Jordan algebra $J_3^{O}$ \cite{Ferrara:1997uz}. 

In the stringy black hole context, $Z_{AB}$ is the central charge 
matrix and $(x,y)$ are the quantized charges of the black hole (28 
electric and 28 magnetic).  The relation between the entropy of stringy black holes and the 
Cartan-Cremmer-Julia $E_{7(7)}$ invariant was established in \cite{Kallosh:1996uy}.  
The central charge 
matrix $Z_{AB}$ can be brought to the canonical basis for the 
skew-symmetric matrix using an $SU(8)$ transformation: 
\be
Z_{ab} =  \pmatrix{ z_1 & 0& 0& 0\cr 0& z_2 &0 &0 \cr
0& 0 & z_3 & 0 \cr 0 & 0 & 0 & z_4 } \otimes \pmatrix{
0& 1 \cr -1 & 0 }  
\ee
where $z_i=\rho_i e^{i\varphi_i}$ are complex. In this way the number 
of entries is reduced from 56 to 8.  In a systematic treatment in 
\cite{Ferrara:1997ci}, the meaning of these parameters was clarified.  
From 4 complex values of  $z_i=\rho_i e^{i\varphi_i}$  
one can remove 3 phases  by an $SU(8)$ rotation, but the overall phase 
cannot be removed; it is related to an extra parameter in the class of 
black hole solutions \cite{Cvetic:1995kv,Cvetic:1995bj}.  In this 
basis, 
the quartic invariant takes the form \cite{Kallosh:1996uy}
\[
I_4 = \sum_i |z_i|^4 -2 \sum_{i<j} |z_i|^2 |z_j|^2 + 4(z_1z_2z_3z_4 +
\bar z_1 \bar z_2 \bar z_3 \bar z_4 )
\]
\[  
 = (\rho_1 + \rho_2 + \rho_3 + \rho_4)
(\rho_1 + \rho_2 - \rho_3 - \rho_4  )
(\rho_1 - \rho_2 + \rho_3 -  \rho_4  )
(\rho_1  - \rho_2 -   \rho_3  + \rho_4  )
\]
\be 
+8 \rho_1 \rho_2 \rho_3 \rho_4 ( \cos  \varphi  -1 ) 
\la{phase}
\ee
Therefore a 5-parameter solution is called a generating solution for 
other black holes in N=8 supergravity/M-theory. The expression for their entropy is 
always given by   
 \be
 S = \pi \sqrt{|I_{4}|}
 \la{entropy}
 \ee
for some subset of 5 of the 8 parameters mentioned 
above. Recently a new class of solutions was discovered, describing 
black rings. The maximal number of parameters for the known 
solutions  is 7. The entropy of black ring solutions found so far was 
identified in \cite{Bena:2004tk, Bena} with the expression 
(\ref{entropy}) for a  subset of 7 out of 8 parameters mentioned 
above. 

Kallosh and Linde have shown that $I_{4}$ depending on 4 complex eigenvalues can be 
represented as Cayley's hyperdeterminant of a hypermatrix $a_{ABD}$. 
To see this, we that in $x, y$ basis only the $SO(8)$ symmetry is manifest, 
which means that every term in  (\ref{cartan}) is invariant only under $SO(8)$ 
symmetry. However, it was proved in \cite{Cartan} and  
\cite{Cremmer:1979up} that the sum of all terms in  (\ref{cartan}) 
is invariant under the full $SU(8)$ symmetry, which acts as follows
\be
\delta(x^{ab} \pm i y_{ab})= (2\Lambda^{[ a}{}_{[c} 
\delta^{b]}{}_{d]} \pm i\Sigma_{abcd}) (x^{cd} \mp i y_{cd}) \ .
\ee
The total number of parameters is 63, where 28 are from the manifest 
$SO(8)$ and 35 from the antisymmetric self-dual $\Sigma_{abcd}= 
{}^*\Sigma^{abcd} $. Thus one can use the $SU(8)$ transformation of 
the complex matrix $x^{ab} + i y_{ab}$ and bring it to the canonical 
form with some complex eigenvalues $\lambda_I,  I=1,2,3,4$. The value 
of the quartic invariant (\ref{cartan}) will not change. 
\be
(x^{ab} + i y_{ab})_{\rm can}= 
\left(
\begin{array}{cccccccc}
0 & \lambda_{1} & 0 & 0 & 0 & 0 & 0 & 0 \\
-\lambda_{1} & 0 & 0 & 0 & 0 & 0 & 0 & 0 \\
0 & 0 & 0 & \lambda_{2} & 0 & 0 & 0 & 0 \\
0 & 0 & -\lambda_{2} &0 & 0 & 0 & 0 & 0 \\
0 & 0 & 0 & 0 & 0 & \lambda_{3} & 0 & 0 \\
0 & 0 & 0 & 0 & -\lambda_{3} & 0 & 0 & 0 \\
0 & 0 & 0 & 0 & 0 & 0 & 0 & \lambda_{4} \\
0 & 0 & 0 & 0 & 0 & 0 & -\lambda_{4} & 0
\end{array}
\right)
\label{Zcan}
\ee
The relation between the complex coefficients $\lambda_{I}$, 
the parameters $x^{ij}$ and $y_{kl}$, the matrix $a_{ABD}$ and the 
black hole charges $p^{i}$ and $q_{k}$ \cite{Duff:2006uz} is given by the following 
dictionary:
\bea\label{dictionary5}
\lambda_{1}~ =& x^{12}+i y_{12}~ =& a_{111}+i a_{000}~ = -q_0-i p^0\nonumber\\
\lambda_{2}~ =& x^{34}+i y_{34}~ =& a_{001}+i a_{110}~ =-p^1+iq_1  \nonumber\\
\lambda_{3}~ =& x^{56}+i y_{56}~ =& a_{010}+i a_{101}~ =-p^2+iq_2  \nonumber\\
\lambda_{4}~ =& x^{78}+i y_{78}~ =& a_{100}+i a_{011}~ =~p^3-iq_3 
\la{charges1}
\end{eqnarray}

If we now write the quartic $E_{7(7)}$ Cartan invariant in the canonical basis $(x^{ij},y_{ij})$, $i,j 
=1,...,8$:
\bea\label{quarticshort}
 I_4 &= -(x^{12}y_{12} + x^{34}y_{34}
+x^{56}y_{56}+x^{78}y_{78})^2-
4(x^{12}x^{34}x^{56}x^{78}+y_{12}y_{34}y_{56}y_{78})\cr & +
4(x^{12}x^{34}y_{12}y_{34}+ x^{12}x^{56}y_{12}y_{56} +
x^{34}x^{56}y_{34}y_{56}+x^{12}x^{78}y_{12}y_{78}+ x^{34}x^{78}
y_{34}y_{78}\cr &+x^{56}x^{78}y_{56}y_{78})\ . 
\eea
then it may now be compared to Cayley's hyperdeterminant 
(\ref{hyperdet}). We find
\be
I_{4}=-{\rm Det}~a
\ee 

The above discussion of $E_{7(7)}$ also applies, mutatis mutandis, to $E_{7}(C)$.

To understand better the entanglement we note that, as a result of 
(\ref{decomp}), Cartan's invariant contains not one Cayley 
hyperdeterminant but seven! It may be written as the sum of seven 
terms each of which is invariant under $[SL(2)]^{3}$ plus cross 
terms.  To see this, denote a $2$ in one of the seven 
entries in (\ref{decomp}) by A, B, C, D, E, F, G. 
So we may rewrite (\ref{decomp}) as
\be
56=(ABD)+(BCE)+(CDF)+(DEG)+(EFA)+(FGB)+(GAC)
\la{561}
\ee
or symbolically
\be
56= a+b+c+d+e+f+g
\la{562}
\ee
Then $I_{4}$ is the singlet in $56 \times 56 \times 56 \times 56$:
\[
J_{4}\sim a^{4}+b^{4}+c^{4}+d^{4}+e^{4}+f^{4}+g^{4}+
\]
\[
2[a^{2}b^{2}+b^{2}c^{2}+c^{2}d^{2}+d^{2}e^{2}+e^{2}f^{2}+f^{2}g^{2}+g^{2}a^{2}+
\]
\[
  a^{2}c^{2}+b^{2}d^{2}+c^{2}e^{2}+d^{2}f^{2}+e^{2}g^{2}+f^{2}a^{2}+g^{2}b^{2}+
\]
\[
 a^{2}d^{2}+b^{2}e^{2}+c^{2}f^{2}+d^{2}g^{2}+e^{2}a^{2}+f^{2}b^{2}+g^{2}c^{2}]
\]
\be  
+8[bcdf+cdeg+defa+efgb+fgac+gabd+abce]
\la{564} 
\ee   
where products like
\[
a^{4}= (ABD)(ABD)(ABD)(ABD)
\]
\be
=\epsilon^{A_{1}A_{2}}\epsilon^{B_{1}B_{2}}\epsilon^{D_{1}D_{4}}\epsilon^{A_{3}A_{4}}\epsilon^{B_{3}B_{4}}\epsilon^{D_{2}D _{3}}
{a}_{A_{1}B_{1}D_{1}}{a}_{A_{2}B_{2}D_{2}}{a}_{A_{3}B_{3}D_{3}}{a}_{A_{4}B_{4}D_{4}}
\ee
exclude four individuals (here Charlie, Emma, Fred and George), products like
\[
a^{2}f^{2}=(ABD)(ABD)(FGB)(FGB)
\]
\be
=\epsilon^{A_{1}A_{2}}\epsilon^{B_{1}B_{2}}\epsilon^{D_{1}D_{4}}\epsilon^{F_{3}F_{4}}\epsilon^{G_{3}G_{4}}\epsilon^{D_{2}B_{3}}
{a}_{A_{1}B_{1}D_{1}}{a}_{A_{2}B_{2}D_{2}}{f}_{F_{3}G_{3}B_{3}}{f}_{F_{4}G_{4}B_{4}}
\ee
exclude two individuals (here Charlie and Emma), and products like
\[
abce= (ABD)(BCE)(CDF)(EFA)
\]
\be
=\epsilon^{A_{1}A_{4}}\epsilon^{B_{1}B_{2}}\epsilon^{C_{2}C_{3}}\epsilon^{D_{1}D_{3}}\epsilon^{E_{2}E_{4}}\epsilon^{F_{3}F_{4}}
{a}_{A_{1}B_{1}D_{1}}{b}_{B_{2}C_{2}E_{2}}{c}_{C_{3}D_{3}F_{3}}{e}_{E_{4}F_{4}A_{4}}
\ee
exclude one individual (here George).

\section{The black hole analogy}

In the STU stringy black hole context 
\cite{Duff:2006uz,Duff:1995sm,Behrndt:1996hu,Kallosh:2006zs} the $a_{ABC}$ 
are integers (corresponding to quantized charges) and hence the symmetry group 
is $[SL(2,Z)]^{3}$ rather than $[SL(2,C)]^{3}$. However, as discussed 
by Levay \cite{Levay:2006kf}, there is a branch of quantum information theory which 
concerns itself with real qubits, called {\it rebits}, for which the 
$a_{ABC}$ are real. (One difference remains, however: one may normalize 
the wave function, whereas for black holes there is no such restriction on 
the charges $a_{ABC}$.) It turns out that there are three reality classes 
which can be characterized by the hyperdeterminant
\[
1)~{\rm Det}~a < 0  
\]
\[
2)~{\rm Det}~a=0  
\]
\be
3)~{\rm Det}~a > 0   
\ee

Case (1) corresponds to the non-separable or GHZ class \cite{Greenberger}, for example,
\be
|\Psi\rangle= \frac{1}{{2}}(-|000\rangle +|011\rangle+|101\rangle +|110\rangle)
\la{GHZ}
\ee
Case (2) corresponds to the separable (A-B-C, A-BC, B-CA, C-AB) and W 
classes,  for example
\be
|\Psi\rangle=\frac{1}{\sqrt{3}}(|100\rangle+|010\rangle+|001\rangle)
\ee
In the string/supergravity interpretation \cite{Duff:2006uz}, cases (1) and (2) were 
shown to correspond to BPS black holes, for which half of the 
supersymmetry is preserved.  Case (1) has non-zero horizon area 
and entropy (``large'' black holes), and case (2) to vanishing 
horizon area and entropy (``small'' black holes), at least at the 
semi-classical level. However, small black holes may acquire a non-zero 
entropy through higher order quantum effects. This entropy also has a quantum 
information interpretation involving bipartite entanglement of the three 
qubits \cite{Kallosh:2006zs}. 

Case (3) is also GHZ, for example the above GHZ state (\ref{GHZ}) with a sign flip 
\be
|\Psi\rangle= \frac{1}{{2}}(|000\rangle +|011\rangle+|101\rangle +|110\rangle)
\ee
In the string/supergravity interpretation, case (3) corresponds to non-BPS black 
holes \cite{Kallosh:2006zs}. With four non-zero charges 
$(q_{0},p^{1},p^{2},p^{3})$ in (\ref{charges1}), for example, an extreme but 
non-BPS black hole \cite{Duff:1994jr} may be obtained by flipping the sign 
\cite{Duff:1996qp} of one of the charges. The canonical GHZ state
\be
|\Psi\rangle= \frac{1}{\sqrt{2}}|111\rangle +\frac{1}{\sqrt{2}}|000\rangle
\ee
also belongs to case (3).

In the $N=8$ theory, ``large'' and ``small'' black holes are classified by 
the sign of $I_{4}$:
\[
1)~I_{4} > 0
\]
\[
2)~I_{4}=0 
\]
\be
3)~I_{4} < 0 
\ee
Once again, non-zero $I_{4}$ corresponds to large black holes, which are BPS for 
$I_{4} > 0$ and non-BPS for $I_{4}< 0$, and vanishing  $I_{4}$ to small 
black holes. However, in contrast to $N=2$, case (1) requires that only 1/8 
of the supersymmetry is preserved, while we may have 1/8, 1/4 or 1/2 for 
case (2).

It is worth noting that the charge orbits corresponding to non-zero 
$I_{4}$ are associated with the following cosets:
\be
\frac{E_{7(7)}}{E_{6(2)}}
\ee
and
\be
\frac{E_{7(7)}}{E_{6(6)}}
\ee
The large black hole solutions can be found \cite{Ferrara:2006em} by 
solving the $N=8$ classical attractor equations \cite{Ferrara:1995ih} when at the attractor 
value the $Z_{AB}$ matrix, in normal form, becomes
\be
Z_{AB}= 
\left(
\begin{array}{cccc}
Z\epsilon & 0 & 0 & 0  \\
0 & 0 & 0 & 0\\
0 & 0 & 0 & 0 \\
0 & 0 & 0 &  0 
\end{array}
\right)
\label{Z1}
\ee
for positive $I_{4}$ and
\be
Z_{AB}=e^{i\pi/4}|Z| 
\left(
\begin{array}{cccc}
\epsilon & 0 & 0 & 0  \\
0 & \epsilon & 0 & 0\\
0 & 0 & \epsilon & 0 \\
0 & 0 & 0 &  \epsilon 
\end{array}
\right)
\label{Z2}
\ee
for negative $I_{4}$. These values exhibit the maximal compact 
symmetries $SU(6) \times SU(2)$ and $USp(8)$ for the positive and 
negative $I_{4}$, respectively.

If the phase in (\ref{phase}) vanishes (which is the case if the configuration
preserves at least 1/4 supersymmetry \cite{Ferrara:1997ci}), $I_4$ becomes
\be
I_4 = \lambda_1 \lambda_2 \lambda_3 \lambda_4~,
\ee
where we have defined $\lambda  _i$ by 
\[
\lambda_1 = \rho_1 + \rho_2 + \rho_3 + \rho_4
\]
\[
\lambda_2 = \rho_1 + \rho_2 - \rho_3 - \rho_4
\]
\[
\lambda_3 = \rho_1 - \rho_2 + \rho_3 - \rho_4 
\]
\be
\lambda_4 = \rho_1 - \rho_2 - \rho_3 + \rho_4
\ee
and we order the $\lambda_i$ so that $\lambda_1 \ge \lambda_2 \ge \lambda_3 \ge |\lambda_4| $. The charge orbits for the small black holes depend on the number of unbroken 
supersymmetries  or the number of vanishing eigenvalues. The orbit is 
\cite{Ferrara:1997uz,Ferrara:1997ci,Lu:1997bg}
\be
\frac{E_{7(7)}}{H_{1,2,3}}
\ee
where
\[
H_{1}=F_{4(4)} \ltimes T_{26}~~~~~~~~\lambda_1,\lambda_2,\lambda_3  \neq 0,~~ \lambda_4 =0~~~(1/8~BPS)
\]
\[
H_{2}=SO(5,6) \ltimes (T_{32} \times T_{1}) ~~~~~~~~\lambda_1,\lambda_2 \neq 0,~~ 
\lambda_3, \lambda_4 =0~~~(1/4~BPS)
\]
\be
H_{3}=E_{6(6)} \ltimes T_{27}~~~~~~~~~ \lambda_1 \neq 
0,~~ \lambda_2 , \lambda_3, \lambda_4=0~~~(1/2~BPS)
\ee 

For $N=8$, as for $N=2$, the large black holes correspond to the two classes of 
GHZ-type (entangled) states and small black holes to the separable or 
W class.

\section{Subsectors}

Having understood the analogy between $N=8$ black holes and the
tripartite entanglement of 7 qubits using $E_{7(7)}$, we may now find the analogy in 
the $N=4$ case using $SL(2) \times  SO(6,6)$ and the $N=2$ case using 
$SL(2) \times SO(2,2) $. 

For $N=4$, as may be seen from (\ref{triality}), we still have an $[SL(2)]^{7}$ subgroup 
but now there are only 24 states
\be
|\Psi\rangle = a_{ABD}|ABD\rangle+e_{EFA}|EFA\rangle+g_{GAC}|GAC\rangle
\ee
So only Alice talks to all the others. This is described by just 
those three lines passing through A in the Fano plane. Then the equations analagous 
to (\ref{561}) and (\ref{562}) are
\be
(2,12)= (ABD)+(EFA)+(GAC)= a+e+g   
\ee        
and the corresponding quartic invariant, $I_{4}$, reduces to the singlet 
in $(2,12) \times(2,12) \times (2,12) \times (2,12) $.
\be
I_{4} \sim a^{4}+e^{4}+g^{4}+2[e^{2}g^{2}+g^{2}a^{2}+a^{2}e^{2}]
\la{2124} 
\ee
If we identify the 24 numbers ($a_{ABD},e_{EFA},g_{GAC}$) with 
$(P^{\mu},Q_{\nu})$ with $\mu,\nu=0,\ldots 11$, this becomes \cite{Duff:1995sm,Cvetic:1995kv,Cvetic:1995bj} 
\be
I_{4}=P^{2}Q^{2}-(P.Q)^{2}
\ee
which is manifestly invarinat under $SL(2) \times  SO(6,6)$.

For $N=2$, as may be seen from (\ref{triality}), we only an $[SL(2)]^{3}$ 
subgroup and there are only 8 states
\be
|\Psi\rangle = a_{ABD}|ABD\rangle
\ee
This is described by just the ABD line in the Fano plane.  This is 
simply the usual tripartite entanglement, for which
\be
(2,2,2)=(ABD)=a
\ee 
and the corresponding quartic invariant
\be
I_{4} \sim a^{4}
\ee
is just Cayley's hyperdeterminant
\be
I_{4}=-{\rm Det} a
\ee

\section{Conclusions}

We note that the 56-dimensional Hilbert space given in (\ref{decomp}) and 
(\ref{psi}) is not a subspace of the usual $2^{7}$-dimensional seven-qubit Hilbert space 
given by $(2,2,2,2,2,2,2)$, but rather a direct sum of seven $2^{3}$-dimensional 
three-qubit Hilbert spaces $(2,2,2)$. This is however, a subspace of 
the $3^{7}$-dimensional seven-qutrit Hilbert space given by 
$(3,3,3,3,3,3,3)$. Under 
\be
 [SL(3)]^{7} \rightarrow [SL(2)]^{7} 
\ee
we have the decomposition
\[
(3,3,3,3,3,3,3) \rightarrow 
\]
\[ 
1~~{\rm term~~like}~~(2,2,2,2,2,2,2)
\]
\[
7~~{\rm terms~~like}~~(2,2,2,2,2,2,1)
\]
\[
21~~{\rm terms~~like}~~(2,2,2,2,2,1,1)
\]
\[
35~~{\rm terms~~like}~~(2,2,2,2,1,1,1)
\]
\[
35~~{\rm terms~~like}~~(2,2,2,1,1,1,1)
\]
\[
21~~{\rm terms~~like}~~(2,2,1,1,1,1,1)
\]
\[
7~~{\rm terms~~like}~~(2,1,1,1,1,1,1)
\]
\be
1~~{\rm term~~like}~~(1,1,1,1,1,1,1)
\ee
which contains
\[
~(2,2,1,2,1,1,1)  
\]
\[ 
+(1,2,2,1,2,1,1)
\]
\[ 
+(1,1,1,2,2,1,2)              
\]
\[
+(2,1,1,1,2,2,1)                                            
\]
\[ 
+(1,2,1,1,1,2,2) 
\]
\be
 +(2,1,2,1,1,1,2)
\ee
So the Fano plane entanglement we have described fits within 
conventional quantum information theory.

The Fano plane also finds application in switching networks that can 
connect any phone to any other phone.  It is the 3-switching network for 
7 numbers.  However there also exists a 4-switching network for 13 numbers, 
a 5-switching network for 21 numbers, and generally an $(n+1)$-switching 
network for  $(n^{2}+n+1)$ numbers corresponding to the projective 
planes of order $n$ \cite{Pegg}. It would be worthwhile pursuing the 
corresponding quantum bit entanglements. 

Exceptional groups, such as $E_{7(7)}$, have featured in 
supergravity, string theory, M-theory and other speculative attempts at 
unification of the fundamental forces. However, it is unusual to find 
an exceptional group appearing in the context of qubit entanglement.
It would be interesting to see whether it can be subject to 
experimental test.

\section{Acknowledgement}

We are grateful to Murat Gunaydin for pointing out the connection between 
our entanglement diagram, Figure 1, and the multiplication table of the 
 octonions. MJD has enjoyed useful conversations with 
Leron Borsten. Help with the diagrams from Laura Andrianopoli is 
also gratefully acknowledged.

This work was supported in part by the National 
Science Foundation under grant number PHY-0245337 and PHY-0555605. 
Any opinions, findings and conclusions or recommendations expressed in this 
material are those of the authors and do not necessarily reflect the views of 
the National Science Foundation. The work of S.F. has been supported in 
part by the European Community Human Potential Program under contract 
MRTN-CT-2004-005104 Ó Constituents, fundamental forces and symmetries 
of the universeÓ, in association with INFN Frascati National Laboratories 
and by the D.O.E grant DE-FG03-91ER40662, Task C. The work of MJD is supported 
in part by PPARC under rolling grant \uppercase{PPA/G/O}/2002/00474, PP/D50744X/1.



\begin{thebibliography}{10}

\bibitem{Duff:2006uz}
 M.~J.~Duff,
``String triality, black hole entropy and Cayley's hyperdeterminant,''
 arXiv:hep-th/0601134.

\bibitem{Kallosh:2006zs}
  R.~Kallosh and A.~Linde,
  ``Strings, black holes, and quantum information,''
  arXiv:hep-th/0602061.

\bibitem{Levay:2006kf}
  P.~Levay,
  ``Stringy black holes and the geometry of entanglement,''
  Phys.\ Rev.\ D {\bf 74}, 024030 (2006)
  [arXiv:hep-th/0603136].

\bibitem{Duff:2006ue}
  M.~J.~Duff and S.~Ferrara,
  ``$E_7$ and the tripartite entanglement of seven qubits,''
  arXiv:quant-ph/0609227.

\bibitem{Levay:2006pt}
  P.~Levay,
   ``Strings, black holes, the tripartite entanglement of seven qubits and the
  Fano plane,''
  arXiv:hep-th/0610314.
 
\bibitem{Duff:1993ij}
  M.~J.~Duff and R.~R.~Khuri,
  ``Four-Dimensional String / String Duality,''
  Nucl.\ Phys.\ B {\bf 411}, 473 (1994)
  [arXiv:hep-th/9305142].

\bibitem{Hull:1994ys}
  C.~M.~Hull and P.~K.~Townsend,
  ``Unity of superstring dualities,''
  Nucl.\ Phys.\ B {\bf 438}, 109 (1995)
  [arXiv:hep-th/9410167].

\bibitem{Duff:1994zt}
  M.~J.~Duff,
  ``Strong / weak coupling duality from the dual string,''
  Nucl.\ Phys.\ B {\bf 442}, 47 (1995)
  [arXiv:hep-th/9501030].

  
\bibitem{Duff:1995sm}
 M.~J.~Duff, J.~T.~Liu and J.~Rahmfeld,
``Four-dimensional string-string-string triality,''
  Nucl.\ Phys.\ B {\bf 459}, 125 (1996)
  [arXiv:hep-th/9508094].

\bibitem{Cayley}
A. Cayley,
``On the theory of linear transformations,''
Camb. Math. J. 4 193-209,1845.

\bibitem{Behrndt:1996hu}
  K.~Behrndt, R.~Kallosh, J.~Rahmfeld, M.~Shmakova and W.~K.~Wong,
  ``STU black holes and string triality,''
  Phys.\ Rev.\ D {\bf 54}, 6293 (1996)
  [arXiv:hep-th/9608059].


\bibitem{Coffman:1999jd}
  V.~Coffman, J.~Kundu and W.~K.~Wootters,
  ``Distributed Entanglement,''
  Phys.\ Rev.\ A {\bf 61}, 052306 (2000)
  [arXiv:quant-ph/9907047].

\bibitem{Miyake}
A. Miyake and M. Wadati,
``Multipartite entanglement and hyperdeterminants,''
[arXiv:quant-ph/0212146].

\bibitem{Duff:2006ev}
  M.~J.~Duff,
  ``Hidden symmetries of the Nambu-Goto action,''
  arXiv:hep-th/0602160.
  
\bibitem{Klein}  
F. Klein,
``Zur Theorie der Liniencomplexe des ersten und zweiten Grades'',
 Math. Ann. 2, 198-226, 1870.
 
 \bibitem{Cartan} 
 E. Cartan, 
 ``Oeuvres completes'', 
 (Editions du Centre National de la Recherche Scientifique, Paris, 1984).
 
 \bibitem{Cremmer:1979up}
  E.~Cremmer and B.~Julia,
  ``The SO(8) Supergravity,''
  Nucl.\ Phys.\ B {\bf 159}, 141 (1979).

 \bibitem{Kallosh:1996uy}
 R.~Kallosh and B.~Kol,
``E(7) Symmetric Area of the Black Hole Horizon,''
 Phys.\ Rev.\ D {\bf 53}, 5344 (1996)
 [arXiv:hep-th/9602014].

 \bibitem{Ferrara:1997uz}
  S.~Ferrara and M.~Gunaydin,
 ``Orbits of exceptional groups, duality and BPS states in string theory,''
  Int.\ J.\ Mod.\ Phys.\ A {\bf 13}, 2075 (1998)
  [arXiv:hep-th/9708025].

\bibitem{Balasubramanian:1997az}
  V.~Balasubramanian, F.~Larsen and R.~G.~Leigh,
  ``Branes at angles and black holes,''
  Phys.\ Rev.\ D {\bf 57}, 3509 (1998)
  [arXiv:hep-th/9704143];
  
 \bibitem{Gunaydin} 
  M.~Gunaydin, K.~Koepsell and H.~Nicolai,
 ``Conformal and quasiconformal realizations of exceptional Lie groups,''
 Commun.\ Math.\ Phys.\  {\bf 221}, 57 (2001)
 [arXiv:hep-th/0008063]. 


\bibitem{Ferrara:1997ci}
  S.~Ferrara and J.~M.~Maldacena,
  ``Branes, central charges and $U$-duality invariant BPS conditions,''
  Class.\ Quant.\ Grav.\  {\bf 15}, 749 (1998)
  [arXiv:hep-th/9706097].
  
\bibitem{Cvetic:1995kv}
  M.~Cvetic and D.~Youm,
   ``All the Static Spherically Symmetric Black Holes of Heterotic String on a
  Six Torus,''
  Nucl.\ Phys.\ B {\bf 472}, 249 (1996)
  [arXiv:hep-th/9512127].
  


\bibitem{Cvetic:1995bj}
  M.~Cvetic and A.~A.~Tseytlin,
  ``Solitonic strings and BPS saturated dyonic black holes,''
  Phys.\ Rev.\ D {\bf 53}, 5619 (1996)
  [Erratum-ibid.\ D {\bf 55}, 3907 (1997)]
  [arXiv:hep-th/9512031].
  
 
\bibitem{Bena:2004tk}
  I.~Bena and P.~Kraus,
  ``Microscopic description of black rings in AdS/CFT,''
  JHEP {\bf 0412}, 070 (2004)
  [arXiv:hep-th/0408186]
  
\bibitem{Bena}  
  I.~Bena, P.~Kraus and N.~P.~Warner,
  ``Black rings in Taub-NUT,''
  Phys.\ Rev.\ D {\bf 72}, 084019 (2005)
  [arXiv:hep-th/0504142].
  
\bibitem{Ferrara:1995ih}
  S.~Ferrara, R.~Kallosh and A.~Strominger,
  ``N=2 extremal black holes,''
  Phys.\ Rev.\ D {\bf 52}, 5412 (1995)
  [arXiv:hep-th/9508072].

\bibitem{Greenberger}
D. M. Greenberger, M. Horne and A. Zeilinger, in {\it Bell's Theorem, 
Quantum Theory and Conceptions of the Universe}, ed. M. Kafatos 
(Kluwer, Dordrecht, 1989)

\bibitem{Duff:1994jr}
  M.~J.~Duff and J.~Rahmfeld,
  ``Massive string states as extreme black holes,''
  Phys.\ Lett.\ B {\bf 345}, 441 (1995)
  [arXiv:hep-th/9406105].
 
\bibitem{Duff:1996qp}
  M.~J.~Duff and J.~Rahmfeld,
  ``Bound States of Black Holes and Other P-branes,''
  Nucl.\ Phys.\ B {\bf 481}, 332 (1996)
  [arXiv:hep-th/9605085].

\bibitem{Ferrara:2006em}
  S.~Ferrara and R.~Kallosh,
  ``On N = 8 attractors,''
  Phys.\ Rev.\ D {\bf 73}, 125005 (2006)
  [arXiv:hep-th/0603247].
    
\bibitem{Lu:1997bg}
  H.~Lu, C.~N.~Pope and K.~S.~Stelle,
  ``Multiplet structures of BPS solitons,''
  Class.\ Quant.\ Grav.\  {\bf 15}, 537 (1998)
  [arXiv:hep-th/9708109].
  

\bibitem{Pegg}
  E.~Pegg~Jr,
  ``The Fano Plane'', MAA online,
  
  Êhttp://www.maa.org/editorial/mathgames/mathgames$\_05\_30\_06$.html

  

   
\bibitem{Ceresole:1994cx}
  A.~Ceresole, R.~D'Auria, S.~Ferrara and A.~Van Proeyen,
  ``On electromagnetic duality in locally supersymmetric N=2 Yang-Mills
  theory,''
  arXiv:hep-th/9412200.

\end{thebibliography}
\end{document}